\documentclass[12pt]{article}
\pdfoutput=1

\usepackage{amsmath,amssymb}
\usepackage{graphicx}

\makeatletter

\renewcommand\section{\@startsection{section}{1}{\z@}
                                   {-3.5ex \@plus -1ex \@minus -.2ex}
                                   {2.3ex \@plus .2ex}
                                   {\normalfont\large\bfseries}}
\renewcommand\subsection{\@startsection{subsection}{2}{\z@}
                                   {-3.25ex\@plus -1ex \@minus -.2ex}
                                   {1.5ex \@plus .2ex}
                                   {\normalfont\normalsize\bfseries}}
\renewcommand\subsubsection{\@startsection{subsubsection}{3}{\z@}
                                   {-3.25ex\@plus -1ex \@minus -.2ex}
                                   {1.5ex \@plus .2ex}
                                   {\normalfont\normalsize\bfseries}}
\renewcommand\paragraph{\@startsection{paragraph}{4}{\z@}
                                   {3.25ex \@plus1ex \@minus.2ex}
                                   {-1em}
                                   {\normalfont\normalsize\bfseries}}

\makeatother

\newcommand{\beq}{\begin{equation}}
\newcommand{\eeq}{\end{equation}}
\newcommand{\bea}{\begin{eqnarray}}
\newcommand{\eea}{\end{eqnarray}}

\newcommand{\SU}{{\rm SU}}

\newcommand{\C}{\mathbb C}

\newcommand{\Z}{\mathbb Z}

\newcommand{\id}{\hbox{1\kern-.27em l}}

\newcommand{\Tr}{{\rm Tr}}

\newcommand{\cC}{{\cal C}}

\newcommand{\cL}{{\cal L}}
\newcommand{\cM}{{\cal M}}

\begin{document}

\pagestyle{empty}

\begin{center}

\vspace*{30mm}
{\LARGE Zero-mode dynamics in supersymmetric Yang-Mills-Chern-Simons theory}

\vspace*{30mm}
{\large M{\aa}ns Henningson}

\vspace*{5mm}
Department of Fundamental Physics\\
Chalmers University of Technology\\
S-412 96 G\"oteborg, Sweden\\[3mm]
{\tt mans@chalmers.se}

\vspace*{30mm}{\bf Abstract:}
\end{center}
We consider minimally supersymmetric Yang-Mills theory with a Chern-Simons term on a flat spatial two-torus in the limit when the torus becomes small. The zero-modes of the fields then decouple from the non-zero modes and give rise to a spectrum of states with energies that are given by multiples of the square of the coupling constant. We discuss the determination of this low-energy spectrum, both for simply connected gauge groups and for gauge groups of adjoint type, with a few examples worked out in detail.

\newpage \pagestyle{plain}

\section{Introduction}
In three space-time dimensions, the Yang-Mills coupling constant $e$ has the dimension of $(\mathrm{mass})^{1/2}$. Such theories are thus trivial in the ultra-violet but can be expected to have non-trivial infrared dynamics. In this paper, we will study them on a flat spatial two-torus
\beq
T^2 = \C / ( L \Z + \tau L \Z) ,
\eeq
where the modular parameter $\tau = \tau_1 + i \tau_2$ is valued in the complex upper half-plane and $L$ is a length (of dimension $(\mathrm{mass})^{-1}$). Most of the quantum states involve the non-zero modes of the fields. Their energy eigenvalues depend on $e$, $L$ and $\tau$  (presumably in a quite complicated way) and diverge in $L \rightarrow 0$ limit. However, some states involve only the zero-modes of the fields, and their energies will be given by finite multiples of $e^2$ in this limit. Computing this weak coupling (or equivalently small volume) spectrum is the aim of the present paper.

More specifically, we will be considering a supersymmetric Yang-Mills-Chern-Simons theory with action
\beq
S = S_{YM} + S_{CS} + S_{Fermion} ,
\eeq
where the usual pure Yang-Mills action
\beq
S_\mathrm{YM} = \frac{1}{4 e^2} \int \Tr \left( F \wedge * F \right)
\eeq
is complemented by a Chern-Simons term \cite{Deser-Jackiw-Templeton}
\beq
S_\mathrm{CS} = \frac{k}{4 \pi} \int \Tr \left(A \wedge d A + \frac{2}{3} A \wedge A \wedge A \right).
\eeq
and fermionic terms
\beq
S_\mathrm{Fermion} = \frac{1}{4 e^2} \int d^3 x \Tr \left( \bar{\lambda} D \!\!\!\! / \lambda \right) + \frac{k}{4 \pi} \int \Tr \left(\bar{\lambda}{\lambda} \right) .
\eeq
Here the fermionic field $\lambda$ is a Majorana spinor in the adjoint representation of the gauge group $G$. For topological reasons, the level $k$ is quantized. This is best expressed in terms of a shifted level \cite{Kao-Lee-Lee}\cite{Amelino-Camelia-Kogan-Szabo}
\beq
k^\prime = k - h / 2 ,
\eeq
where $h$ is the dual Coxeter number of $G$. For a simply connected gauge group
\beq
 k^\prime = 0 \mod \Z , 
\eeq
and for a gauge group of adjoint type (i.e. a simply connected group divided by its center)
\beq
k^\prime = 0 \mod h \Z .
\eeq

This theory has a mass-gap which is visible already in perturbation theory, which makes many general problems of quantum field theory more accessible. By supersymmetry, the energy spectrum is non-negative, and a very basic question is to understand the structure of zero-energy states. The number of such states (i.e. the Witten index of the theory \cite{Witten82}) was elegantly determined for simply connected gauge groups in \cite{Witten99} by algebro-geometric means. In \cite{Henningson}, this computation was redone in a more pedestrian and straightforward fashion. (Similar results were obtained in \cite{Smilga}.) The latter method has the advantage of also allowing for a determination of the actual wave functions of the states. In the cases with a gauge groups of adjoint type, it allows for a refinement where the states can be further classified by their discrete electric and magnetic 't~Hooft fluxes. For convenience, the main results of \cite{Henningson} will be reviewed in the present paper, but we will not repeat all the details. Our main focus here is instead to go further in this direction by determining the complete spectrum of low-energy states (i.e. states the energies of which remain finite in the $L \rightarrow 0$ limit). A natural next step would be to determine the complete spectrum in the weak coupling limit (i.e. the $e \rightarrow 0$ limit with $L$ fixed). We note that somewhat similiar questions for Yang-Mills-Chern-Simons theory in non-compact space have been considered in e.g. \cite{Karabali-Kim-Nair}.

In the next section, we will describe the low energy theory. The determination of the low-energy spectrum for a simply connected gauge group is discussed in section three and performed in full detail in the $\SU (2)$ and $\SU (3)$ cases. In section four, we consider the refinement with discrete electric and magnetic 't Hooft fluxes that is possible with a gauge group of adjoint form \cite{'t_Hooft} and exemplify with $\SU (2) / \Z_2$.

\section{The low-energy theory}
Starting with the bosonic degrees of freedom, we note that the Hamiltonian of the theory contains a magnetic contribution proportional to $\frac{1}{e^2} L^{-2}$ and the square of the magnetic field strength. This term diverges as $L \rightarrow 0$ unless it vanishes, so in this limit, the low energy theory localizes on gauge field configurations of zero magnetic field strength. Such spatially flat connections are completely determined by their commuting holonomies around the two cycles of $T^2$. By a gauge transformation (acting by conjugation) these can be taken to lie in a maximal torus subgroup of $G$. Furthermore, the complex structure of $T^2$ induces a complex structure on the space of such pairs of holonomies. They can thus be assembled into an element $Z$ of the complex torus 
\beq
X = \C \otimes V / (\Gamma \otimes \Lambda) .
\eeq
Here $V$ is the root space and $\Lambda$ the root lattice of the gauge group $G$ so that $V / \Lambda$ is a maximal torus, and $\Gamma$ is the lattice $\Z + \tau \Z$ so that $\C / \Gamma$ is a torus `dual' to the spatial $T^2$. The bosonic low-energy degrees of freedom $Z$ can be viewed as the zero-modes of the gauge field $A$ restricted from the non-abelian Lie algebra of $G$ to a Cartan subalgebra. Finally, we must identify elements of $X$ that are related by the residual gauge symmetry, so the moduli space of spatially flat connections is really
\beq
\cM = X / W ,
\eeq
where $W$ is the Weyl group of $G$.

The vector space underlying the torus $X$ has a metric $\cC_{a b}$, $a, b = 1, \ldots, \mathrm{rank} \; G$ given by the Killing form of $G$ restricted to the maximal torus, and the bosonic variable $Z$ should in more detail be written as a multi component vector $Z^a$. We will suppress this from the notation, though, and just denote the scalar product between such vectors with a raised dot. Hopefully, this will not cause any confusion.

We continue with the fermionic degrees of freedom. At a generic point of the moduli space $\cM$, the gauge group $G$ is spontaneously broken by the holonomies to the abelian subgroup $V$. Some components of the spinor field $\lambda$ are then massive and can be integrated out. This procedure is responsible for the shift of the Chern-Simons level from $k$ to $k^\prime = k - h / 2$ as mentioned in the introduction \cite{Kao-Lee-Lee}\cite{Amelino-Camelia-Kogan-Szabo}. We denote the remaining fermionic degrees of freedom, which consist of spatially constant $V$-valued modes, as $\eta_+$ and $\eta_-$.
\footnote{The subscript indicates the spatial chirality.}

The dynamics of the low-energy degrees of freedom $Z$, $\bar{Z}$, $\eta_+$ and $\eta_-$ is governed by the action
\bea
S & = & \frac{1}{4 e^2} \int d t \left( \frac{d Z}{d t} \cdot \frac{d \bar{Z}}{d t} + \eta_+ \cdot \frac{d \eta_-}{d t} + \eta_- \cdot \frac{d \eta_+}{d t} \right) \cr
& & + \frac{\pi k^\prime}{\tau - \bar{\tau}} \int d t \left(  Z \cdot \frac{d \bar{Z}}{d t} - \bar{Z} \cdot \frac{d Z}{d t} + \eta_+ \cdot \eta_- \right) .
\eea
The conserved supercharges are
\bea
Q_+ & = & \frac{1}{4 e^2} \eta_- \cdot \frac{d Z}{d t}  \cr
Q_- & = & \frac{1}{4 e^2} \eta_+ \cdot \frac{d \bar{Z}}{d t} .
\eea
By a standard canonical analysis and quantization in a Schr\"odinger representation, they correspond to operators
\bea
Q_+ & = & e \eta_- \cdot i \frac{D}{D Z} \cr
Q_- & = & e \eta_+ \cdot i \frac{D}{D \bar{Z}} .
\eea
Here the covariant derivatives are given by
\bea
\frac{D}{D Z} & = & \frac{\partial}{\partial Z} - \frac{i \pi k^\prime}{\tau - \bar{\tau}} \bar{Z} \cr
\frac{D}{D \bar{Z}} & = & \frac{\partial}{\partial \bar{Z}} + \frac{i \pi k^\prime}{\tau - \bar{\tau}} Z
\eea
and fulfil the commutation relations
\bea
\left[ \frac{D}{D Z}, \frac{D}{D Z} \right] & = & 0 \cr
\left[ \frac{D}{D \bar{Z}}, \frac{D}{D \bar{Z}} \right] & = & 0 \cr 
\left[ \frac{D}{D Z}, \frac{D}{D \bar{Z}} \right] & = & \frac{\pi k^\prime}{\tau_2} \cC ,
\eea
i.e. the different components of $\frac{D}{D Z}$ commute with each other, whereas the commutators between the components $\frac{D}{D Z}$ and $\frac{D}{D \bar{Z}}$ are given by a multiple of the Killing metric. Analogously, the fermionic operators fulfil the canonical anti-commutation relations
\bea
\left\{ \eta_+, \eta_+ \right\} & = & 0 \cr
\left\{ \eta_-, \eta_- \right\} & = & 0 \cr
\left\{ \eta_+, \eta_- \right\} & = & \frac{\tau_2}{\pi} \cC .
\eea
As usual in a supersymmetric theory, the Hamiltonian can be written in a manifestly positive definite way as
\bea
H & = & \left\{ Q_+, Q_- \right\} \cr
& = & e^2 \left( \frac{\tau_2}{\pi} i \frac{D}{D Z} \cdot i \frac{D}{D \bar{Z}} - \frac{\pi k^\prime}{\tau_2} \eta_- \cdot \eta_+  \right) .
\eea
The operators $\frac{D}{D Z}$ and $\eta_+$ can now be interpreted as bosonic and fermionic creation operators respectively in the sense that
\bea
\left[ H, \frac{D}{D Z} \right] & = & e^2 k^\prime \frac{D}{D Z} \cr
\left[ H, \eta_+ \right] & = & e^2 k^\prime \eta_+ ,
\eea
i.e. they raise the energy by an amount $e^2 k^\prime$. The corresponding annihilation operators $\frac{D}{D \bar{Z}}$ and $\eta_-$ obey
\bea
\left[ H, \frac{D}{D \bar{Z}} \right] & = & - e^2 k^\prime \frac{D}{D Z} \cr
\left[ H, \eta_- \right] & = & -e^2 k^\prime \eta_- 
\eea
and lower the energy by $e^2 k^\prime$. The supercharges $Q_+$ and $Q_-$ of course commute with the Hamiltonian $H$:
\beq
\left[ H, Q_+ \right] = \left[ H, Q_- \right] = 0 .
\eeq

The bosonic Hilbert space consists of square integrable `wave sections' $\Psi (Z, \bar{Z})$ of a certain line bundle $\cL^{k^\prime}$ over the complex torus $X$ \cite{Henningson}. The fermionic Hilbert space is spanned by states obtained by acting with components of the creation operators $\eta_+$ on a fermionic vacuum state $| 0 \rangle$ which is annihilated by all components of $\eta_-$. The ground states of the theory (which are annihilated by $Q_+$ and $Q_-$ and thus have zero energy) are then precisely given by the states of the form
\beq
\Psi (Z, \bar{Z}) \otimes | 0 \rangle
\eeq
where $\Psi$ obeys the holomorphicity condition 
\beq
\frac{D}{D \bar{Z}} \Psi (Z, \bar{Z}) = 0 .
\eeq
A distinguished solution to this condition is given by
\beq \label{Psi}
\Psi (Z, \bar{Z}) = \exp \left( \frac{i \pi k^\prime}{\tau - \bar{\tau}} (Z - \bar{Z}) \cdot Z \right) \psi (Z)
\eeq
with the holomorphic theta function
\beq
\psi (Z) = \sum_{\lambda \in \Lambda} \exp \left(i \pi k^\prime \tau \lambda \cdot \lambda + 2 \pi i k^\prime \lambda \cdot Z \right) .
\eeq
A basis of solutions is then given by
\beq
(T_{\epsilon \tau} \Psi) (Z, \bar{Z}) = \exp \left( \frac{i \pi k^\prime}{\tau - \bar{\tau}} (Z - \bar{Z}) \cdot Z \right) (T_{\epsilon \tau} \psi) (Z) 
\eeq
for $\epsilon \in \frac{1}{k^\prime} \Lambda^* / \Lambda$. Here $\Lambda^*$ is the weight lattice of $G$ which is dual to the root lattice $\Lambda$, and $T_{\epsilon \tau} \psi$ is obtained by translating $\psi$ on $X$:
\bea
(T_{\epsilon \tau} \psi) (Z) & = & \pm \exp (i \pi k^\prime \tau \epsilon \cdot \epsilon + 2 \pi i k^\prime \epsilon \cdot Z) \psi (Z + \epsilon \tau) \cr
& = & \sum_{\lambda \in \Lambda} \exp \left( i \pi k^\prime \tau (\lambda + \epsilon) \cdot (\lambda + \epsilon) + 2 \pi i k^\prime (\lambda + \epsilon) \cdot Z \right) .
\eea
The prefactor is included to give the quasi periodicity conditions appropriate for a section of $\cL^{k^\prime}$. The Weyl group $W$ of $G$ leaves $\psi$ invariant and acts on the sections $T_{\epsilon \tau} \psi$ via its permutation action on $\epsilon \in \frac{1}{k^\prime} \Lambda^* / \Lambda$. See \cite{Henningson} for more details on this construction.

It appears that the complete set of states can be obtained by acting on the ground states $(T_{\epsilon \tau} \Psi) (Z, \bar{Z}) \otimes | 0 \rangle$ with the bosonic and fermionic creation operators $\frac{D}{D Z}$ and $\eta_+$. The Weyl group $W$ acts not only on the ground states as described in the previous paragraph, but also on these excitations. The space of gauge invariant physical states is obtained by projecting the complete set of states onto its Weyl invariant subspace. 

\section{Simply connected gauge groups}
Rather than describing the procedure to determine the spectrum in general, we will discuss two concrete examples with gauge groups $G = \SU (2)$ and $G = \SU (3)$ in detail. Other groups should be amenable to an analogous analysis, but the computations would get more involved. We do not expect any striking new phenomena to occur, though.

\subsection{$G = \SU (2)$}
With gauge group $G = \SU (2)$, the Weyl group $W$ is isomorphic to the permutation group $S_2$ on two objects. It has two irreducible unitary representations, both of dimension one: The trivial representation ${\bf 1}$, and another representation ${\bf 1^\prime}$ in which the non-trivial element acts as $-1$. The tensor products of these are of course given by the multiplication table
\beq
\begin{tabular}{rll}
\vline & ${\bf 1}$ & ${\bf 1^\prime}$ \cr
\hline
${\bf 1}$ \vline & ${\bf 1}$ & ${\bf 1^\prime}$ \cr
${\bf 1^\prime}$ \vline & ${\bf 1^\prime}$ & ${\bf 1}$ .
\end{tabular}
\eeq

The ground state label $\epsilon$ takes its values in
\beq
\frac{1}{k^\prime} \Lambda^* / \Lambda \simeq \left\{ (a, b) \in \left( \frac{1}{2 k^\prime} \Z / \Z \right) \times \left( \frac{1}{2 k^\prime} \Z / \Z \right) \Big| a + b = 0 \right\} .
\eeq
The Weyl group $W$ acts by permutation on $(a, b)$, so there are two Weyl invariants
\bea
& & \Bigl(0, 0 \Bigr) \cr
& & \left( \frac{1}{2}, \frac{1}{2} \right) 
\eea
and $k^\prime - 1$ Weyl pairs
\beq
\left( \frac{l}{2 k^\prime}, \frac{2 k^\prime - l}{2 k^\prime} \right), \left( \frac{2 k^\prime - l}{2 k^\prime}, \frac{l}{2 k^\prime} \right)\;\;\; \mathrm{for} \;\;\; l = 1, \ldots, k^\prime - 1 .
\eeq
This gives $k^\prime + 1$ ground states that transform in the ${\bf 1}$ representation
\beq
\Psi_{\bf 1} = \left\{
\begin{array}{l}
\Psi \otimes | 0 \rangle \cr
T_{\left( \frac{1}{2}, \frac{1}{2} \right) \tau} \Psi \otimes | 0 \rangle \cr
\frac{1}{\sqrt{2}} \left( T_{\left( \frac{l}{2 k^\prime}, \frac{2 k^\prime - l}{2 k^\prime} \right) \tau} + T_{\left( \frac{2 k^\prime - l}{2 k^\prime}, \frac{l}{2 k^\prime} \right) \tau} \right) \Psi \otimes | 0 \rangle \;\;\; \mathrm{for} \;\;\; l = 1, \ldots, k^\prime - 1 .
\end{array}
\right.
\eeq
and $k^\prime - 1$ ground states in the ${\bf 1^\prime}$ representation
\beq
\Psi_{\bf 1^\prime} = \frac{1}{\sqrt{2}} \left( T_{\left( \frac{l}{2 k^\prime}, \frac{2 k^\prime - l}{2 k^\prime} \right) \tau} - T_{\left( \frac{2 k^\prime - l}{2 k^\prime}, \frac{l}{2 k^\prime} \right) \tau} \right) \Psi \otimes | 0 \rangle \;\;\; \mathrm{for} \;\;\; l = 1, \ldots, k^\prime - 1 .
\eeq
(Here $\Psi$ is the distinguished Weyl invariant solution (\ref{Psi}).) All these ground states have bosonic statistics, and the gauge invariant states are those that transform in the ${\bf 1}$ representation, so we get the physical ground state spectrum
\bea
k^\prime + 1 & \mathrm{bosonic \;\; states \;\; with} & E = 0 .
\eea

The bosonic and fermionic creation operators $\frac{D}{D Z}$ and $\eta_+$ both transform in the ${\bf 1^\prime}$ representation under $W$. Gauge invariant excited states can now be obtained in two different ways: Either by acting on a ground state in the ${\bf 1}$ representation with an even number of creation operators
\bea
\frac{D^{2 n}}{D Z^{2 n}} \Psi_{\bf 1} \otimes | 0 \rangle & & \mathrm{bosonic}\cr
\frac{D^{2 n - 1}}{D Z^{2 n - 1}} \Psi_{\bf 1} \otimes \eta_+ | 0 \rangle & & \mathrm{fermionic} ,
\eea
or by acting on a ground state in the ${\bf 1^\prime}$ representation with an odd number of operators
\bea
\frac{D^{2 n + 1}}{D Z^{2 n + 1}} \Psi_{\bf 1^\prime} \otimes | 0 \rangle & & \mathrm{bosonic}\cr
\frac{D^{2 n}}{D Z^{2 n}} \Psi_{\bf 1^\prime} \otimes \eta_+ | 0 \rangle & & \mathrm{fermionic} .
\eea
So the spectrum of excited states is
\bea
k^\prime + 1 & \mathrm{bosonic \;\; and \;\; fermionic \;\; states \;\; with} & E = 2 n e^2 k^\prime \cr
k^\prime - 1 & \mathrm{bosonic \;\; and \;\; fermionic \;\; states \;\; with} & E = (2 n + 1) e^2 k^\prime .
\eea
The second formula is valid for $n = 0, 1, \ldots$, and the first formula only for $n = 1, 2, \ldots$. 

\subsection{$G = \SU (3)$}
With gauge group $G = \SU (3)$, the Weyl group $W$ is isomorphic to the permutation group $S_3$ on three objects. It has three irreducible unitary representations ${\bf 1}$, ${\bf 1^\prime}$, and ${\bf 2}$ of the indicated dimensions. The tensor products of these are given by the multiplication table
\beq
\begin{tabular}{rlll}
\vline & ${\bf 1}$ & ${\bf 1^\prime}$ & ${\bf 2}$ \cr
\hline
${\bf 1}$ \vline & ${\bf 1}$ & ${\bf 1^\prime}$ & ${\bf 2}$ \cr
${\bf 1^\prime}$ \vline & ${\bf 1^\prime}$ & ${\bf 1}$ & ${\bf 2}$ \cr
${\bf 2}$ \vline & ${\bf 2}$ & ${\bf 2}$ & $\left[ {\bf 1} \oplus {\bf 2} \right]_S \oplus \left[ {\bf 1^\prime} \right]_A$ .
\end{tabular}
\eeq

The ground state label $\epsilon$ takes its values in
\beq
\frac{1}{k^\prime} \Lambda^* / \Lambda \simeq \left\{ (a, b, c) \in \left( \frac{1}{3 k^\prime} \Z / \Z \right) \times \left( \frac{1}{3 k^\prime} \Z / \Z  \right) \times \left( \frac{1}{3 k^\prime} \Z / \Z \right) \Big| a + b + c = 0 \right\} .
\eeq
The Weyl group acts by permutation on $(a, b, c)$, so there are three Weyl invariants
\beq
\left( \frac{l}{3 k^\prime}, \frac{l}{3 k^\prime}, \frac{l}{3 k^\prime} \right) \;\;\; \mathrm{for} \;\;\; l = 0, k^\prime, 2 k^\prime
\eeq
$3 k^\prime - 3$ Weyl triplets
\beq
\left( \frac{l}{3 k^\prime}, \frac{l}{3 k^\prime}, \frac{3 k^\prime - 2 l}{3 k^\prime} \right), \left( \frac{l}{3 k^\prime}, \frac{3 k^\prime - 2 l}{3 k^\prime}, \frac{l}{3 k^\prime} \right), \left( \frac{3 k^\prime - 2 l}{3 k^\prime}, \frac{l}{3 k^\prime}, \frac{l}{3 k^\prime} \right) \;\;\; \mathrm{for} \;\;\; l \neq 0, k^\prime, 2 k^\prime 
\eeq
and $\frac{1}{2} (k^\prime - 1) (k^\prime -2)$ Weyl sextets
\bea
(a, b, c), (b, c, a), (c, a, b), (c, b, a), (b, a, c), (a, c, b) 
\eea
with $a, b, c \in \frac{1}{3 k^\prime} \Z / \Z$ all different but $a = b = c \mod \frac{1}{k^\prime}$. The Weyl invariant elements give rise to ground states in the ${\bf 1}$ representation
\beq
\Psi_{\bf 1} = T_{\left( \frac{l}{3 k^\prime}, \frac{l}{3 k^\prime}, \frac{l}{3 k^\prime} \right) \tau} \Psi \otimes | 0 \rangle .
\eeq
Each Weyl triplet gives rise to a ${\bf 1}$ state
\beq
\Psi_{\bf 1} = \frac{1}{\sqrt{3}} \left( T_{\left( \frac{l}{3 k^\prime}, \frac{l}{3 k^\prime}, \frac{3 k^\prime - 2 l}{3 k^\prime} \right) \tau} + T_{\left( \frac{l}{3 k^\prime}, \frac{3 k^\prime - 2 l}{3 k^\prime}, \frac{l}{3 k^\prime} \right) \tau} + T_{\left( \frac{3 k^\prime - 2 l}{3 k^\prime}, \frac{l}{3 k^\prime}, \frac{l}{3 k^\prime} \right) \tau} \right) \Psi \otimes | 0 \rangle
\eeq
and two states in the ${\bf 2}$ representation
\beq
\Psi_{\bf 2} = \left\{ \begin{array}{l}
\frac{1}{\sqrt{3}} \left( T_{\left( \frac{l}{3 k^\prime}, \frac{l}{3 k^\prime}, \frac{3 k^\prime - 2 l}{3 k^\prime} \right) \tau} + e^{2 \pi i / 3} T_{\left( \frac{l}{3 k^\prime}, \frac{3 k^\prime - 2 l}{3 k^\prime}, \frac{l}{3 k^\prime} \right) \tau} + e^{4 \pi i / 3} T_{\left( \frac{3 k^\prime - 2 l}{3 k^\prime}, \frac{l}{3 k^\prime}, \frac{l}{3 k^\prime} \right) \tau} \right) \Psi \otimes | 0 \rangle \cr
\frac{1}{\sqrt{3}} \left( T_{\left( \frac{l}{3 k^\prime}, \frac{l}{3 k^\prime}, \frac{3 k^\prime - 2 l}{3 k^\prime} \right) \tau} + e^{4 \pi i / 3} T_{\left( \frac{l}{3 k^\prime}, \frac{3 k^\prime - 2 l}{3 k^\prime}, \frac{l}{3 k^\prime} \right) \tau} + e^{2 \pi i / 3} T_{\left( \frac{3 k^\prime - 2 l}{3 k^\prime}, \frac{l}{3 k^\prime}, \frac{l}{3 k^\prime} \right) \tau} \right) \Psi \otimes | 0 \rangle 
\end{array} \right.
\eeq
Each Weyl sextet gives rise to a ${\bf 1}$ state
\beq
\Psi_{\bf 1} = \frac{1}{\sqrt{6}} \left( T_{(a, b, c) \tau} + T_{(b, c, a) \tau} + T_{(c, a, b) \tau} + T_{(c, b, a) \tau} + T_{(b, a, c) \tau} + T_{(a, c, b) \tau} \right)  \Psi \otimes | 0 \rangle ,
\eeq
a ${\bf 1^\prime}$ state
\beq
\Psi_{\bf 1^\prime} = \frac{1}{\sqrt{6}} \left( T_{(a, b, c) \tau} + T_{(b, c, a) \tau} + T_{(c, a, b) \tau} - T_{(c, b, a) \tau} - T_{(b, a, c) \tau} - T_{(a, c, b) \tau} \right)  \Psi \otimes | 0 \rangle ,
\eeq
and four states transforming as two copies of ${\bf 2}$
\beq
\Psi_{\bf 2} = \left\{ \begin{array}{l} 
\frac{1}{\sqrt{3}} \left( T_{(a, b, c) \tau} + e^{2 \pi i / 3} T_{(b, c, a) \tau} + e^{4 \pi i / 3} T_{(c, a, b) \tau} \right) \Psi \otimes | 0 \rangle \cr
\frac{1}{\sqrt{3}} \left( T_{(c, b, a) \tau} + e^{2 \pi i / 3} T_{(b, a, c) \tau} + e^{4 \pi i / 3} T_{(a, c, b) \tau} \right) \Psi \otimes | 0 \rangle 
\end{array} \right.
\eeq
and
\beq
\Psi_{\bf 2} = \left\{ \begin{array}{l} 
\frac{1}{\sqrt{3}} \left( T_{(a, b, c) \tau} + e^{4 \pi i / 3} T_{(b, c, a) \tau} + e^{2 \pi i / 3} T_{(c, a, b) \tau} \right) \Psi \otimes | 0 \rangle \cr
\frac{1}{\sqrt{3}} \left( T_{(c, b, a) \tau} + e^{4 \pi i / 3} T_{(b, a, c) \tau} + e^{2 \pi i / 3} T_{(a, c, b) \tau} \right) \Psi \otimes | 0 \rangle 
\end{array} \right. .
\eeq
Altogether, the ground states transform in the representation
\beq
R_\mathrm{ground \; states} = \frac{1}{2} (k^\prime + 2) (k^\prime + 1) \times {\bf 1} \oplus \frac{1}{2} (k^\prime - 1) (k^\prime - 2) \times {\bf 1^\prime} \oplus (k^{\prime 2} - 1) \times {\bf 2} .
\eeq
(By this we mean a direct sum of $\frac{1}{2} (k^\prime + 2) (k^\prime + 1)$ copies of ${\bf 1}$, $\frac{1}{2} (k^\prime - 1) (k^\prime - 2)$ copies of ${\bf 1^\prime}$, and $k^{\prime 2} - 1$ copies of ${\bf 2}$.) All these ground states have bosonic statistics, and the gauge invariant states are those in the ${\bf 1}$ representation, so we get the physical ground state spectrum
\bea
\frac{1}{2} (k^\prime + 2) (k^\prime + 1) & \mathrm{bosonic \;\; states \;\; with} & E = 0 .
\eea

The bosonic and fermionic creation operators $\frac{D}{D Z}$ and $\eta_+$ both transform in the ${\bf 2}$ representation under $W$. Multiple bosonic excitations give rise to the symmetric product representations
\bea
\left[ {\bf 2} \right]^{\otimes 6 n}_S & = & (n + 1) \times {\bf 1} \oplus n \times {\bf 1^\prime} \oplus 2 n \times {\bf 2}  \cr
\left[ {\bf 2} \right]^{\otimes (6 n + 1)}_S & = & n \times {\bf 1} \oplus n \times {\bf 1^\prime} \oplus (2 n + 1) \times {\bf 2}  \cr
\left[ {\bf 2} \right]^{\otimes (6 n + 2)}_S & = & (n + 1) \times {\bf 1} \oplus n \times {\bf 1^\prime} \oplus (2 n + 1) \times {\bf 2}  \cr
\left[ {\bf 2} \right]^{\otimes (6 n + 3)}_S & = & (n + 1) \times {\bf 1} \oplus (n + 1)\times {\bf 1^\prime} \oplus (2 n + 1) \times {\bf 2}  \cr
\left[ {\bf 2} \right]^{\otimes (6 n + 4)}_S & = & (n + 1) \times {\bf 1} \oplus n \times {\bf 1^\prime} \oplus (2 n + 2) \times {\bf 2}  \cr
\left[ {\bf 2} \right]^{\otimes (6 n + 5)}_S & = & (n + 1) \times {\bf 1} \oplus (n + 1) \times {\bf 1^\prime} \oplus (2 n + 2) \times {\bf 2}
\eea
whereas zero, one or two fermionic excitations give the anti-symmetric products
\bea
\left[ {\bf 2} \right]^{\otimes 0}_A & = & {\bf 1} \cr
\left[ {\bf 2} \right]^{\otimes 1}_A & = & {\bf 2} \cr
\left[ {\bf 2} \right]^{\otimes 2}_A & = & {\bf 1^\prime} .
\eea
At energy level $E = m e^2 k^\prime$ for $m = 1, 2, \ldots$, we have bosonic excitations transforming in the representation
\beq
R_m = \left[ {\bf 2} \right]^{\otimes m}_S \otimes \left[ {\bf 2} \right]^{\otimes 0}_A \oplus \left[ {\bf 2} \right]^{\otimes (m - 2)}_S \otimes \left[ {\bf 2} \right]^{\otimes 2}_A .
\eeq
By supersymmetry, this has to agree with the fermionic representation
\beq
R_m = \left[ {\bf 2} \right]^{\otimes (m - 1)}_S \otimes \left[ {\bf 2} \right]^{\otimes 1}_A .
\eeq
It is given by
\bea
R_{3 n} & = & n \times {\bf 1} \oplus n \times {\bf 1^\prime} \oplus 2n \times {\bf 2} \cr
R_{3 n + 1} & = & n \times {\bf 1} \oplus n \times {\bf 1^\prime} \oplus (2n + 1) \times {\bf 2} \cr
R_{3 n + 2} & = & (n + 1) \times {\bf 1} \oplus (n + 1) \times {\bf 1^\prime} \oplus (2n + 1) \times {\bf 2} .
\eea
Gauge invariant excited states correspond to ${\bf 1}$ terms in the tensor product of these excitation representations with the ground state representation $R_\mathrm{ground \; state}$. The spectrum of excited states is thus
\bea
3 k^{\prime 2} n & \mathrm{bosonic \;\; and \;\; fermionic \;\; states \;\; with} & E = 3 n e^2 k^\prime \cr
3 k^{\prime 2} n + k^{\prime 2} - 1 & \mathrm{bosonic \;\; and \;\; fermionic \;\; states \;\; with} & E = (3 n + 1) e^2 k^\prime \cr
3 k^{\prime 2} n + 2 k^{\prime 2} + 1 & \mathrm{bosonic \;\; and \;\; fermionic \;\; states \;\; with} & E = (3 n + 2) e^2 k^\prime .
\eea
The second and third formulas are valid for $n = 0, 1, \ldots$, and the first formula only for $n = 1, 2, \ldots$. 

\section{Gauge groups of adjoint type}
A gauge group of the form $G = \hat{G} / C$, where $\hat{G}$ is a simply connected group with center subgroup $C = \Lambda^* / \Lambda$, is said to be of adjoint type. It imposes the requirement $k^\prime = 0 \mod h \Z$ on the shifted level $k^\prime$, where $h$ is the dual Coxeter number of $G$. The theory is then invariant under translations $T_{\mu + \nu \tau}$ on the complex torus $X$ with $\mu, \nu \in \Lambda^* / \Lambda$. Physical states can be characterized by the phase factor $\exp 2 \pi i (\mu \cdot e_1 + \nu \cdot e_2)$ picked up under such a transformation. This defines the discrete electric 't~Hooft fluxes $e_1, e_2 \in \Lambda^* / \Lambda$. In our case, these transformations act on the ground states as
\beq \label{tHooft-shift}
\left( T_{\mu + \nu \tau} T_{\epsilon \tau} \Psi \right) (Z) = \exp \left( i \pi k^\prime \mu \cdot \mu + i \pi k^\prime \nu \cdot \nu + 2 \pi i k^\prime \epsilon \cdot \mu \right) \left( T_{(\nu + \epsilon) \tau} \Psi \right) (Z) 
\eeq
and commute with the creation operators $\frac{D}{D Z}$ and $\eta_+$. The phase factor in the transformation law of the ground states thus determines $e_1$, whereas state of definite $e_2$ are obtained by taking appropriate linear combinations of the states $T_{(\nu + \epsilon) \tau} \Psi$ for $\nu \in \Lambda^* / \Lambda$. There is also a magnetic 't~Hooft flux $m_{12}$ which measures the obstruction to lifting the structure group of the gauge bundle from $G = \hat{G} / C$ to its universal covering group $\hat{G}$.

Here we will only consider the simplest case of $G = \SU (2) / \Z_2$. The case of $G = \SU (3) / \Z_3$ can be easily worked out by extending the ground state analysis of section 3.2 of the present paper and section 3.2 of \cite{Henningson}. Other groups would involve more work.

\subsection{$G = \SU (2) / \Z_2$}
The non-trivial element $\left( \frac{1}{2}, \frac{1}{2} \right)$ of the center subgroup $C = \Lambda^* / \Lambda \simeq \Z_2$, has norm $\left( \frac{1}{2}, \frac{1}{2} \right) \cdot \left( \frac{1}{2}, \frac{1}{2} \right) = \frac{1}{2}$. The dual Coxeter number of $\SU (2)$ is $h = 2$, so the shifted level $k^\prime$ has to be an even integer. In fact, it is convenient to consider the two cases when $k^\prime$ is divisible by four or not separately.

\subsubsection{$k^\prime = 0 \mod 4$}
The transformation law (\ref{tHooft-shift}) now simplifies to
\beq \label{simpletHooft}
\left( T_{\mu + \nu \tau} T_{\epsilon \tau} \Psi \right) (Z) = \exp \left( 2 \pi i k^\prime \epsilon \cdot \mu \right) \left( T_{(\nu + \epsilon) \tau} \Psi \right) (Z) .
\eeq

The ground states in the ${\bf 1}$ representation can be further classified by their electric 't~Hooft flux components $e_1$ and $e_2$ which determine the phases under translations by $\left( \frac{1}{2}, \frac{1}{2} \right)$ and $\left( \frac{1}{2}, \frac{1}{2} \right) \tau$ on $X$. There are $k^\prime / 4 + 1$ states with both $e_1$ and $e_2$ trivial:
\beq
\Psi_{\bf 1} = \left\{
\begin{array}{l}
\frac{1}{\sqrt{2}} \left( T_{(0, 0) \tau} + T_{\left( \frac{1}{2}, - \frac{1}{2} \right) \tau} \right) \Psi \cr
\frac{1}{\sqrt{2}} \left( T_{\left( \frac{1}{4},  - \frac{1}{4} \right) \tau} + T_{\left( \frac{3}{4}, - \frac{3}{4} \right) \tau} \right) \Psi \cr
\frac{1}{2} \left( T_{\left( \frac{l}{2 k^\prime}, -  \frac{l}{2 k^\prime} \right) \tau} + T_{\left( \frac{2 k^\prime - l}{2 k^\prime}, -  \frac{2 k^\prime - l}{2 k^\prime} \right) \tau} + T_{\left( \frac{k^\prime + l}{2 k^\prime}, -  \frac{k^\prime + l}{2 k^\prime} \right) \tau} +  T_{\left( \frac{k^\prime - l}{2 k^\prime}, -  \frac{k^\prime - l}{2 k^\prime} \right) \tau} \right) \Psi ,
\end{array}
\right.
\eeq
where $l = 2, 4, \ldots, k^\prime / 2 - 2$ in the last line. There are also $k^\prime / 4$ states for each remaining combination of $e_1$ and $e_2$:
\beq
\Psi_{\bf 1} = \left\{
\begin{array}{l}
\frac{1}{\sqrt{2}} \left( T_{(0, 0) \tau} - T_{\left( \frac{1}{2}, - \frac{1}{2} \right) \tau} \right) \Psi \cr
\frac{1}{2} \left( T_{\left( \frac{l}{2 k^\prime}, -  \frac{l}{2 k^\prime} \right) \tau} + T_{\left( \frac{2 k^\prime - l}{2 k^\prime}, -  \frac{2 k^\prime - l}{2 k^\prime} \right) \tau} - T_{\left( \frac{k^\prime + l}{2 k^\prime}, -  \frac{k^\prime + l}{2 k^\prime} \right) \tau} -  T_{\left( \frac{k^\prime - l}{2 k^\prime}, -  \frac{k^\prime - l}{2 k^\prime} \right) \tau} \right) \Psi
\end{array}
\right. ,
\eeq
where $l = 2, 4, \ldots, k^\prime / 2 - 2$ in the last line, have $e_1$ trivial and $e_2$ non-trivial.
\beq
\Psi_{\bf 1} = \frac{1}{2} \left( T_{\left( \frac{l}{2 k^\prime}, -  \frac{l}{2 k^\prime} \right) \tau} + T_{\left( \frac{2 k^\prime - l}{2 k^\prime}, -  \frac{2 k^\prime - l}{2 k^\prime} \right) \tau} + T_{\left( \frac{k^\prime + l}{2 k^\prime}, -  \frac{k^\prime + l}{2 k^\prime} \right) \tau} +  T_{\left( \frac{k^\prime - l}{2 k^\prime}, -  \frac{k^\prime - l}{2 k^\prime} \right) \tau} \right) \Psi ,
\eeq
where $l = 1, 3, \ldots, k^\prime / 2 - 1$, have $e_1$ non-trivial and $e_2$ trivial.
\beq
\Psi_{\bf 1} = \frac{1}{2} \left( T_{\left( \frac{l}{2 k^\prime}, -  \frac{l}{2 k^\prime} \right) \tau} + T_{\left( \frac{2 k^\prime - l}{2 k^\prime}, -  \frac{2 k^\prime - l}{2 k^\prime} \right) \tau} - T_{\left( \frac{k^\prime + l}{2 k^\prime}, -  \frac{k^\prime + l}{2 k^\prime} \right) \tau} -  T_{\left( \frac{k^\prime - l}{2 k^\prime}, -  \frac{k^\prime - l}{2 k^\prime} \right) \tau} \right) \Psi ,
\eeq
where $l = 1, 3, \ldots, k^\prime / 2 - 1$, have both $e_1$ and $e_2$ non-trivial. Acting on these states with an even number of creation operators gives excited states as described in the previous section.

The ground states in the ${\bf 1^\prime}$ representation can be treated analogously. There are $k^\prime / 4 - 1$ states with both $e_1$ and $e_2$ trivial:
\beq
\Psi_{\bf 1^\prime} = \frac{1}{2} \left( T_{\left( \frac{l}{2 k^\prime}, -  \frac{l}{2 k^\prime} \right) \tau} - T_{\left( \frac{2 k^\prime - l}{2 k^\prime}, -  \frac{2 k^\prime - l}{2 k^\prime} \right) \tau} + T_{\left( \frac{k^\prime + l}{2 k^\prime}, -  \frac{k^\prime + l}{2 k^\prime} \right) \tau} -  T_{\left( \frac{k^\prime - l}{2 k^\prime}, -  \frac{k^\prime - l}{2 k^\prime} \right) \tau} \right) \Psi ,
\eeq
where $l = 2, 4, \ldots, k^\prime / 2 - 2$. There are also $k^\prime / 4$ states for each remaining combination of $e_1$ and $e_2$:
\beq
\Psi_{\bf 1^\prime} = \left\{ \begin{array}{l}
\frac{1}{\sqrt{2}} \left( T_{\left( \frac{1}{4},  - \frac{1}{4} \right) \tau} - T_{\left( \frac{3}{4}, - \frac{3}{4} \right) \tau} \right) \Psi \cr
\frac{1}{2} \left( T_{\left( \frac{l}{2 k^\prime}, -  \frac{l}{2 k^\prime} \right) \tau} - T_{\left( \frac{2 k^\prime - l}{2 k^\prime}, -  \frac{2 k^\prime - l}{2 k^\prime} \right) \tau} - T_{\left( \frac{k^\prime + l}{2 k^\prime}, -  \frac{k^\prime + l}{2 k^\prime} \right) \tau} +  T_{\left( \frac{k^\prime - l}{2 k^\prime}, -  \frac{k^\prime - l}{2 k^\prime} \right) \tau} \right) \Psi ,
\end{array}
\right. ,
\eeq
where $l = 2, 4, \ldots, k^\prime / 2 - 2$ in the last line, have $e_1$ trivial and $e_2$ non-trivial. 
\beq
\Psi_{\bf 1^\prime} = \frac{1}{2} \left( T_{\left( \frac{l}{2 k^\prime}, -  \frac{l}{2 k^\prime} \right) \tau} - T_{\left( \frac{2 k^\prime - l}{2 k^\prime}, -  \frac{2 k^\prime - l}{2 k^\prime} \right) \tau} + T_{\left( \frac{k^\prime + l}{2 k^\prime}, -  \frac{k^\prime + l}{2 k^\prime} \right) \tau} -  T_{\left( \frac{k^\prime - l}{2 k^\prime}, -  \frac{k^\prime - l}{2 k^\prime} \right) \tau} \right) \Psi ,
\eeq
where $l = 1, 3, \ldots, k^\prime / 2 - 1$, have $e_1$ non-trivial and $e_2$ trivial.
\beq
\Psi_{\bf 1^\prime} = \frac{1}{2} \left( T_{\left( \frac{l}{2 k^\prime}, -  \frac{l}{2 k^\prime} \right) \tau} - T_{\left( \frac{2 k^\prime - l}{2 k^\prime}, -  \frac{2 k^\prime - l}{2 k^\prime} \right) \tau} - T_{\left( \frac{k^\prime + l}{2 k^\prime}, -  \frac{k^\prime + l}{2 k^\prime} \right) \tau} +  T_{\left( \frac{k^\prime - l}{2 k^\prime}, -  \frac{k^\prime - l}{2 k^\prime} \right) \tau} \right) \Psi ,
\eeq
where $l = 1, 3, \ldots, k^\prime / 2 - 1$, have both $e_1$ and $e_2$ non-trivial. These states should be acted on by an odd number of creation operators to obtain physical excited states.

The states described above all have trivial magnetic 't~Hooft flux, i.e. the gauge bundle can be lifted to an $\SU (2)$ bundle over $T^2$. There is also an additional gauge-invariant ground state with non-trivial magnetic 't~Hooft flux. Like the other ground states, this state has bosonic statistics \cite{Henningson}. There are however no excitations with energies of order $e^2$ above this ground state.

\subsubsection{$k^\prime = 2 \mod 4$}
The transformation law now acquires additional signs for non-trivial $\mu$ or $\nu$ as compared to (\ref{simpletHooft}). The analysis is otherwise precisely analogous, and we give only the results for the number of ground states of different quantum numbers:

There are $(k^\prime + 2) / 4 - 1$ ground states in the ${\bf 1}$ representation with both $e_1$ and $e_2$ trivial, and $(k^\prime + 2) / 4$ ground states for each of the tree remaining combinations of $e_1$ and $e_2$.

There are $(k^\prime + 2) / 4$ ground states in the ${\bf 1^\prime}$ representation with both $e_1$ and $e_2$ trivial, and $(k^\prime + 2) / 4 - 1$ ground states for each of the tree remaining combinations of $e_1$ and $e_2$.

Excited states are obtained by acting on these ground states with an even or odd number of creation operators respectively.

Finally, there is a single ground state with non-trivial magnetic 't~Hooft flux. An interesting feature of this state is that it has fermionic statistics \cite{Henningson}. There are no low-energy excitations above this ground state either.

\vspace*{3mm}
This research was supported by the Swedish Research Council.

\end{document}